\documentclass[preprint,authoryear]{elsarticle}

\usepackage{sectsty}
\usepackage{graphicx}
\usepackage{amsmath}
\usepackage{amssymb}
\usepackage{multicol}
\usepackage{changes}
\usepackage{float}
\usepackage{booktabs}
\usepackage{subcaption}
\usepackage[font={small,it}]{caption}
\usepackage{multirow}
\usepackage{comment}
\usepackage{natbib}
\journal{Spatial Statistics}
\makeatletter
\newcommand*\bigcdot{\mathpalette\bigcdot@{.5}}
\newcommand*\bigcdot@[2]{\mathbin{\vcenter{\hbox{\scalebox{#2}{$\m@th#1\bullet$}}}}}

\excludecomment{mycomment}
\makeatother

\begin{document}
	\begin{frontmatter}
\title{Accounting for spatial varying sampling effort due to accessibility in Citizen Science data: A case study of moose in Norway}

\author[1]{Jorge Sicacha-Parada\corref{cor1}%
}
\ead{jorge.sicacha@ntnu.no}
\author[1]{Ingelin Steinsland}
\ead{ingelin.steinsland@ntnu.no}
\author[3]{Benjamin Cretois}
\ead{bernjamin.cretois@ntnu.no}
\author[4]{Jan Borgelt}
\ead{jan.borgelt@ntnu.no}

\cortext[cor1]{Corresponding author}
\address[1]{Department of Mathematical Sciences. NTNU (Norwegian University of Science and Technology), Norway)}
 \address[3]{Department of Geography. NTNU, Norway}
 \address[4]{Department of Energy and Process Engineering. NTNU, Norway}

	%\pagebreak
	
	% Optional TOC
	% \tableofcontents
	% \pagebreak
	
	%--Paper--
	%--------------- Section 1 ---------------
\begin{abstract}
    Citizen Scientists together with an increasing access to technology provide large datasets that can be used to study e.g. ecology and biodiversity. Unknown and varying sampling effort is a major issue when making inference based on citizen science data.
    In this paper we propose a modeling approach for accounting for variation in sampling effort due to accessibility. The paper is based on a illustrative case study using citizen science data of moose occurrence in Hedmark, Norway. The aim is to make inference about the importance of two geographical properties known to influence moose occurrence; terrain ruggedness index and solar radiation. Explanatory analysis show that moose occurrences are overrepresented close to roads, and we use distance to roads as a proxy for accessibility. 
 We propose a model based on a Bayesian Log-Gaussian Cox Process specification for occurrence. The model accounts for accessibility through a distance sampling approach. This approach can be seen as a thinning process where probability of thinning, i.e. not observing, increases with increasing distances. For the moose case study distance to roads are used.  Computationally efficient full Bayesian inference is performed using the Integrated Nested Laplace Approximation and the Stochastic Partial Differential Equation approach for spatial modeling.  The proposed model as well as the consequences of not accounting for varying sampling effort due to accessibility are studied through a simulation study based on the case study. Considerable biases are found in estimates for the effect of radiation on moose occurrence when accessibility is not considered in the model.

\end{abstract}
\end{frontmatter}

	\section{Introduction}

With the expansion of technology, information and data have become readily available not only for the scientific community, but also for society in general. Citizen Science (CS), i.e. the engagement of the public in activities formerly exclusive of trained people in scientific projects, has emerged as a consequence, \citep{doi:10.1890/110294}. The convenience offered by technology has encouraged people to contribute to different fields of scientific research ranging from social sciences (www.ancientlives.org, www.oldweather.org) or astronomy (www.galaxyzoo.org) to biodiversity (e.g. www.artsobservasjoner.no, www.eBird.org and www.iNaturalist.org).\\ \\
According to the typology of Citizen Science introduced in \citet{strasser2019citizen}, CS projects in biodiversity are regarded as "sensing" projects. It means that the role of volunteers is to collect information and submit it to a large database. These projects take advantage of the participants local knowledge on their environment and reach high spatial coverage. The impact of these projects can be measured in the amount of observations that are stored in their databases. For example, by September 2019, about 1.3 billion of occurrences had been reported in the global biodiversity information facility (GBIF). The Norwegian biodiversity information centre (Artsdatabanken) has about 21 million of occurrences reported. Despite being cost-efficient, easy to retrieve and its massive amount, CS data have some drawbacks. Given their "open" nature, there is no systematic sampling design to collect data, meaning citizens record observations at convenient sampling locations and times. Additionally, no scientific background is required to be part of a CS project, which implies that some species may get misidentified, \citep{kelling2015can}.\\ \\
The differences in knowledge and expertise of participants in CS projects is only one of the potential sources of bias. As described in \citet{doi:10.1111/2041-210X.12254}, the biases in the sampling processes can be classified in four groups: temporal bias, understood as varying activity of observation and reporting across time; geographical bias, meaning more reports in more convenient locations, \citep{10.1371/journal.pone.0147796}; uneven sampling effort per visit and differences in detectability. Preference for reporting a specific type of species constitutes another typical bias in CS sampling designs. All these biases yield in uneven sampling effort across space and time. Moreover the sampling process is not always independent of the variable intended to be measured or observed, known as preferential sampling, \citep{doi:10.1111/j.1467-9876.2009.00701.x}. An issue that is not exclusive to CS records and that needs to be considered when uncertain about the independence between observation and sampling design.\\ \\
Furthermore, ideally citizens record both locations where species have been observed and locations where species have been absent. This type of data is known as presence-absence data. In this case the locations are fixed and presence or absence of a species is recorded. However, CS databases in biodiversity contain mostly presence-only data. Hence, the only information given is the presence of a species in random locations whereas the rest of the landscape remains unknown. They can be actual absences or locations that have not been sampled yet. Then, there is an evident necessity of modeling CS data in a way that acknowledges the randomness of the number and the location of the observations and that accounts for different biases in the underlying sampling process.\\ \\
The focus of this paper is on presence-only data and geographical bias due to accessibility. A common approach to model this data is turning some of the unobserved locations into pseudo-absences, then the available observations could be modeled as presence-absence data, \citep{Ferrier2002,doi:10.1111/j.2041-210X.2011.00172.x} However, it does not account for the spatial autocorrelation for presences and absences across space, \citep{doi:10.1002/ecm.1372}. Arguably the most common approach for modeling presence-only data is Maxent,\citep{doi:10.1890/07-2153.1,PHILLIPS2006231} . This is an algorithmic strategy that aims to find an optimal species density subject to some constraint. Given its nature, Maxent does not account for the uncertainty of the predictions. Furthermore, it provides the relative chance of finding a species in comparison to other locations rather than a probability of presence or absence at each location. In \citet{doi:10.1111/j.1467-9876.2011.00769.x} presence-only data is regarded as a realization of a spatial point process which for the particular case of CS data is subject to degradation. This approach was proven to perform better than Maxent in terms of goodness-of-fit statistics in a scenario with biased sampling.\\ \\
The source of variation in sampling effort targeted in this paper is spatial bias due to differences in accessibility. It has been discussed in \citet{doi:10.1002/ecm.1372} and addressed in \citet{doi:10.1111/ecog.03944} that studies historical large mammal records in South Africa where accessibility depends on proximity to freshwater and European settlements. There, an accessibility index is computed as the average of two functions defined as the half-normal function, characteristic of distance sampling. This functional form is also mentioned in \citet{yuan2017point} as an approach to model the probability of detection as a function of the perpendicular distance to a transect line segment.\\ \\
In this paper we aim to emphasize the importance of accounting for differences in accessibility when CS data is modeled. We do it by making use of the Bayesian spatial approach proposed in \citet{doi:10.1111/j.1467-9876.2011.00769.x} and \citet{doi:10.1002/ecm.1372} to model the intensity of the point process associated to the distribution of a species. It means the observed point process is understood as the resulting process after the potential point process has been degraded by the probability of having access to each location. Our working hypothesis is that the distance to the road system is a good indicator of accessibility. Thus, we account for accessibility by making use of the functional form presented in \citet{yuan2017point}, which links the distance to the closest road to the probability of accessing the location. This functional form is then included as part of the model that explains the observed intensity. We refer to this model as the varying sampling effort model. A common goal of ecological studies is to explore the importance of  geographical, climatic or biological quantities that drive the distribution of a species. Hence, we also aim to see how accounting for accessibility impacts the parameters estimates in a Bayesian spatial model, changing then the way the dynamics of a species is understood.  \citet{doi:10.1002/ecm.1372} uses a Markov chain Monte Carlo (MCMC) sampling for inference, which is computationally expensive. The Integrated Nested Laplace Approximation (INLA), \citep{doi:10.1111/j.1467-9868.2008.00700.x} is non-sampling approach to full Bayesian inference. INLA can also be used for spatial models based on Gaussian Matern Processes using the stochastic partial differential equation (SPDE) approach, \citep{doi:10.1111/j.1467-9868.2011.00777.x} , also in point process modeling, \citep{10.1093/biomet/asv064}. We use INLA for inference, and the computational efficiency enable us to do a simulation study.\\ \\
In this paper we consider an illustrative case study of CS presence data of moose (\textit{Alces alces}) in the county of Hedmark, Norway. Moose is a large ungulate distributed across most of the Norwegian landscape. It utilizes a wide variety of environments, including forests, wetlands and farmland, \citep{Hundertmark2016}. The species contributes to ecosystem health parameters by providing key ecological processes such as browsing on both broad-leaved and needle-leaved trees as well as shrubs (for a review see \citet{shipley2010fifty}). Moose survival and fitness is highly determined by competition for food, e.g. \citet{messier1991significance}. Hence, moose tend to avoid areas dominated by steep slopes, deep and enduring snow cover as well as poor food availability. In order to proxy this knowledge , we use two explanatory variables: solar radiation (RAD) and terrain ruggedness index (TRI). Solar radiation has been shown to influence fine scale movement of moose due to its effects on air temperature, snow cover and plant phenology, \citep{doi:10.1002/(SICI)1099-1085(199812)12:15<2339::AID-HYP800>3.0.CO;2-L}. Moose are more likely to select areas receiving higher levels of solar energy as snow cover is shallow and plant productivity higher. Ruggedness, or terrain heterogeneity also has a major role in moose distribution as a high ruggedness increase their energy expenditure, \citep{doi:10.1111/j.1600-0587.2009.06104.x}.\\ \\
This paper is organized as follows: In Section 2, the dataset of the case study is introduced and explored. In Section 3, models are presented, as well as the inference method and measures for evaluating and comparing models. In Section 4, a simulation study comparing the varying sampling model and a model not accounting for varying sampling effort. In Section 5 results of both the simulation study and the moose case study are shown. The paper finishes in Section 6 with the discussion of the results and concluding remarks.

\section{Case study: Moose in Hedmark and Exploratory Analysis}
In this paper we study moose distribution using locations recorded by citizen scientists and retrieved from GBIF (https://gbif.org). It corresponds to 472 observations product of human observation from 2000 to 2019. These observations correspond to locations of moose in the county of Hedmark, Norway, see Figure 1a. Further, we have two explanatory variables available: RAD and TRI. RAD is computed as the yearly average of the monthly solar radiation retrieved from WorldClim (http://worldclim.org/version2), \citep{doi:10.1002/joc.5086}. TRI was obtained from the ENVIREM dataset (https://envirem.github.io). Both variables are available at approximately 1km $\times$ 1km resolution, \citep{doi:10.1111/ecog.02880}. \\ \\
Our working hypothesis is that spatial variation in sampling effort can be partly explained by accessibility due to distance to roads. In order to determine whether or not it happens, we used the road system of Hedmark retrieved from the spatial crowd-sourcing project OpenStreetMap (https://www.openstreetmap.org). This dataset includes a detailed network of roads that ranges from highways to footways. Figure 1a shows the roads as well as reported moose presences in Hedmark. Most of the observations are made in southern Hedmark and near populated zones of the region, such as Hamar, Elverum and Kongsvinger, or in zones with many roads.\\
\begin{figure}
\center
	\includegraphics[width=0.4\textwidth]{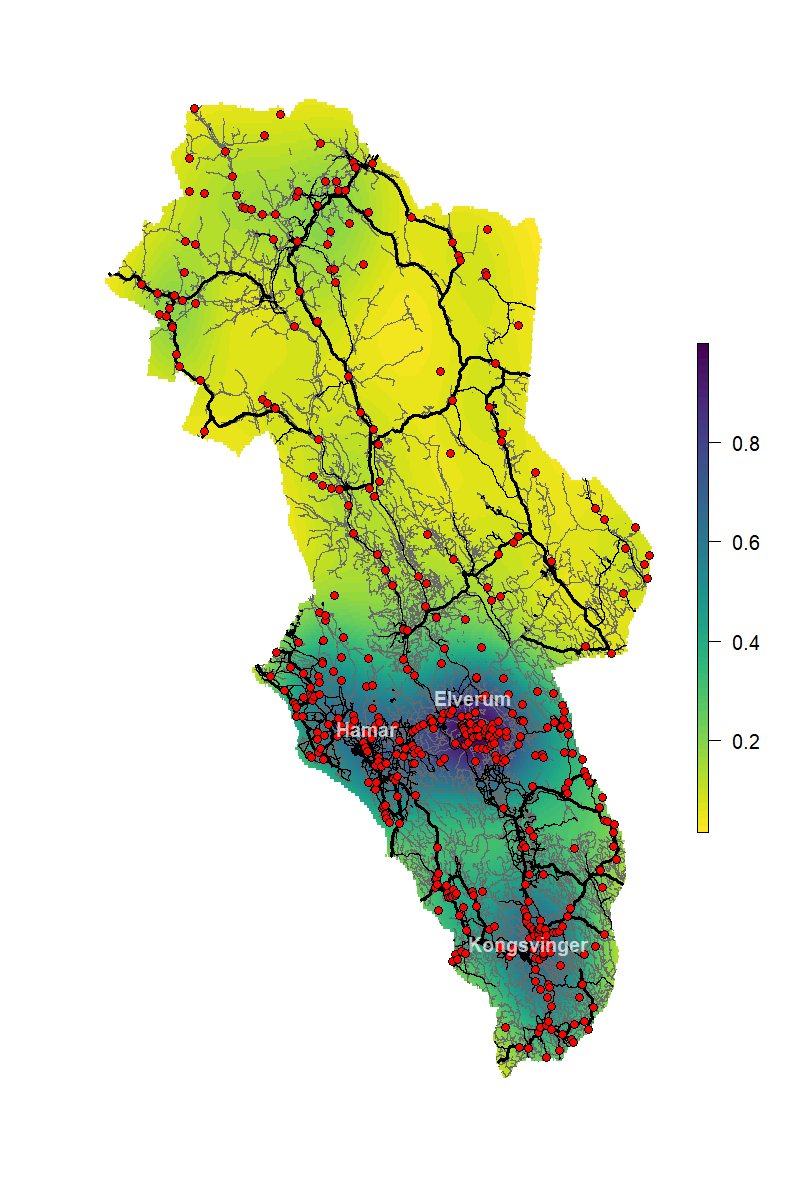} 
	\caption{Moose observations (red points) and road system (lines) in the county of Hedmark, Norway. Bold lines indicate main roads.}
\end{figure}

To explore if the observed locations are more accessible than the mass of locations in the region, we compare the citizen science dataset that contains the 472 observed points with a grid of about 400 thousand evenly distributed points. We computed the closest distance to the road network for both datasets.  The boxplots and empirical cumulative distribution function of the previously mentioned distances for each set of points are displayed in Figure 2.\\
\begin{figure}
		\begin{multicols}{2}
	\includegraphics[width=0.5\textwidth]{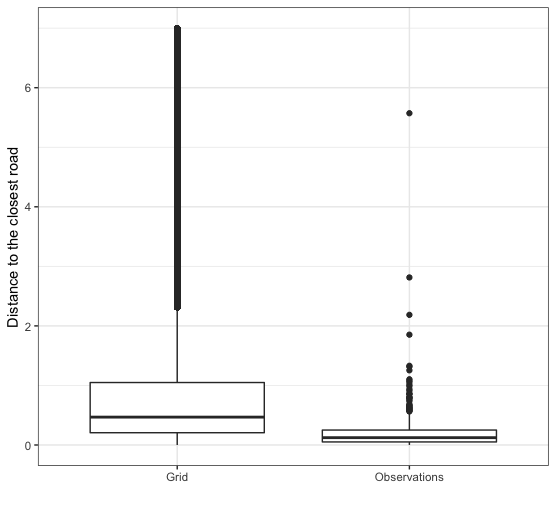} 
\columnbreak
	\includegraphics[width=0.5\textwidth]{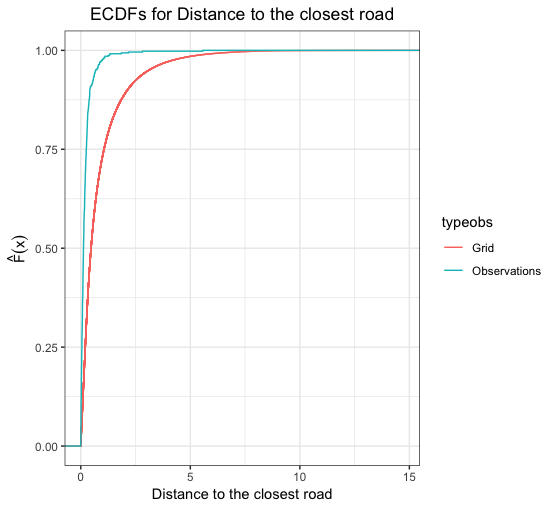} 
	\end{multicols}
	\begin{minipage}{0.5\textwidth}
	\center
\subcaption{ (a)}
\end{minipage}
\begin{minipage}{0.5\textwidth}
\center
\subcaption{(b)}
\end{minipage}
	\caption{ (a) Boxplots of distance to the road system. Left: Dense grid of about 400 thousand points. Right: 472 reports of moose in Hedmark  (b) Empirical Cumulative Distribution Function of distances to the closest road. Red: Dense grid of about 400 thousand points. Blue: 472 reports of moose in Hedmark  }
	\end{figure}
	
91\% of the observations reported are located less than 500 meters away from a road. On the other hand, the grid has points that are more distant from the road system. The boxplots in Figure 2 show that locations further away than 1 kilometer are not represented in the observed point process. These differences are also displayed in the ECDFs plot, which shows how the ECDF of the observed points (blue line) reaches values close to 1 faster than the points in the grid (red line). To confirm these differences, a Kolmogorov-Smirnov test was performed in order to determine if these two sets of distances follow the same distribution or not. The result ($p-value<2.2e-16$) let us conclude that, as suspected, the sets of distances do not follow the same distribution.\\ \\

\begin{figure}[t]
		\begin{multicols}{2}
	\includegraphics[width=0.5\textwidth]{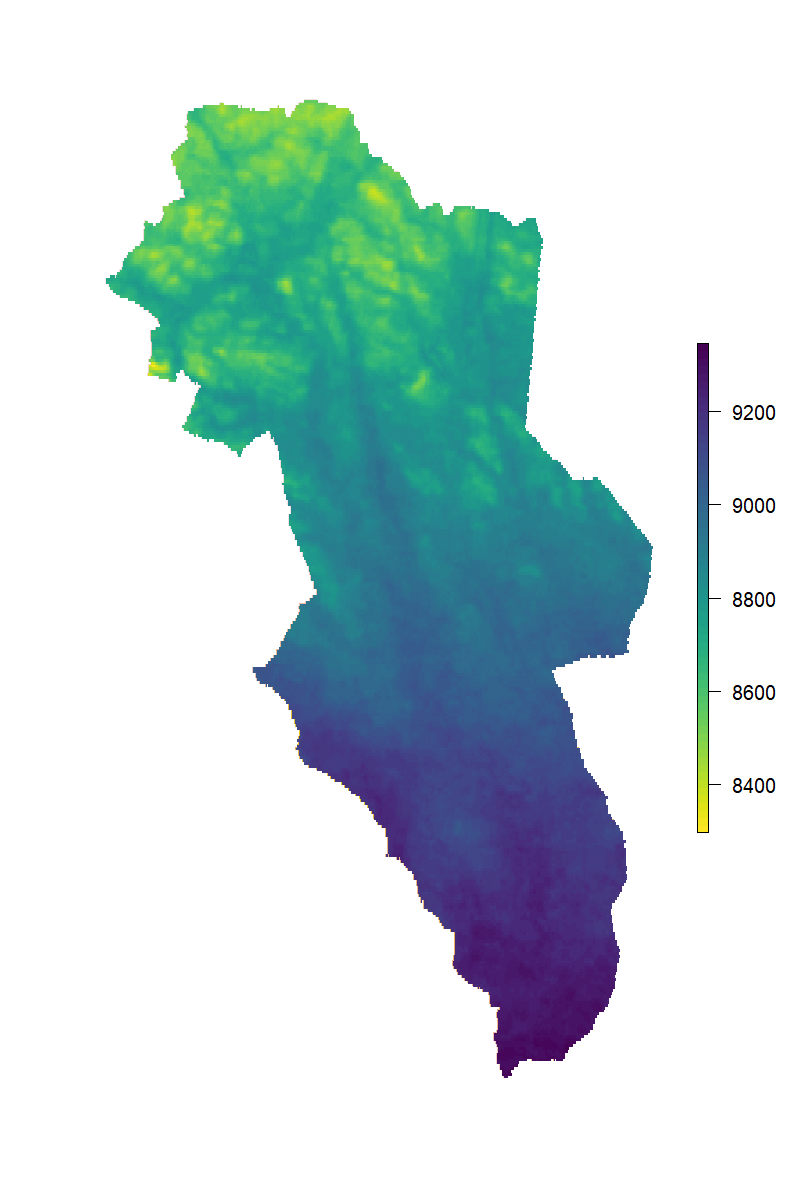} 
\columnbreak
\includegraphics[width=0.5\textwidth]{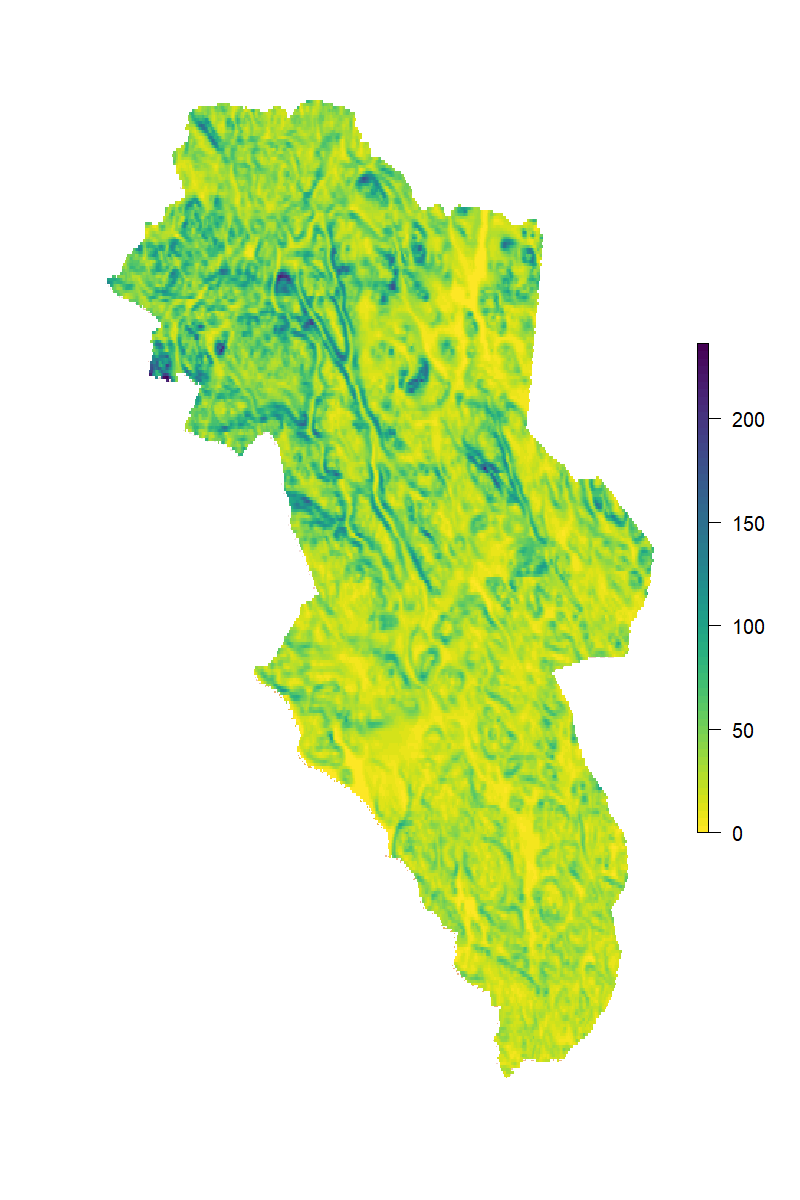} 
	\end{multicols}
		\begin{minipage}{0.5\textwidth}
	\center
\subcaption{ (a)}
\end{minipage}
\begin{minipage}{0.5\textwidth}
\center
\subcaption{(b)}
\end{minipage}
	\caption{ (a) Solar Radiation (RAD) and (b) Terrain Ruggedness Index (TRI) in the county of Hedmark, Norway.}
\end{figure}

\section{Modeling and inference approach}

In this section we introduce two models that will be fitted and compared. They are based on the specification of a Log-Gaussian Cox Process. The first of them, the naive model, does not account for any difference in accessibility, while the second model accounts for accessibility as a potential source of variation in sampling effort (VSE). Then, we briefly describe the inference methods we will use. Finally, we introduce the criteria to assess and compare these models.

\subsection{Models}

\subsubsection{Naive model}
The observed data are regarded as a realization of a point process. It means both the number of points and their locations are random. The intensity measure, understood as the mean number of points per area unit, is the variable we are interested in modeling. In what follows, we will assume the observed point pattern is a realization of an inhomogenous Poisson Process (NHPP), \citep{illian2008statistical}, over the region $D \subset \mathbb{R}^2$. Thus, the number of points in $D$ is assumed to be random and to have a Poisson distribution with mean $\int_D \lambda(x)dx$. We assume the point process is a Log-Gaussian Cox Process (LGCP). Hence, $\lambda(\textbf{s}), \textbf{s} \in D$ can be expressed as:
\begin{equation}
log(\lambda(\textbf{s}))=\textbf{x}^T(\textbf{s})\beta + \omega(\textbf{s})
\label{NaiveModel}
\end{equation}   
with $\textbf{x}(\textbf{s})$ a set of spatially-referenced covariates and $\omega(\textbf{s})$ a zero-mean Gaussian process that accounts for residual spatial autocorrelation between locations in $D$. For our case study the set of spatial covariates $\textbf{x}(\textbf{s})$ are: TRI and RAD, displayed in Figure 3. A flexible family of covariance functions is the Matérn class:
\begin{equation}
\frac{\sigma^2}{\Gamma(\nu)2^{\nu-1}}(\kappa \|s_i-s_j\|)^\nu K_\nu (\kappa\|s_i-s_j\|)
\label{MaternCov}
\end{equation}
with $\|s_i-s_j\|$ the Euclidean distance between two locations $s_i$, $s_j \in D$. $\sigma^2$ stands for the marginal variance, and $K_\nu$ represents the modified Bessel function of the second kind and order $\nu >0$. $\nu$ is the parameter that determines the degree of smoothness of the process, while $\kappa>0$ is a scaling parameter.\\

\subsubsection{VSE model}
Degeneration of the point process has to be considered in the model. We associate it to a thinned intensity. That is, we now assume that the intensity of the observed point process is $\lambda(\textbf{s})q(\textbf{s})$ with $\lambda(\textbf{s})$ the intensity modeled in the naive model, named in \citet{doi:10.1111/j.1467-9876.2011.00769.x} as the potential intensity and $q(\textbf{s})$ the thinning factor which ranges between 0 and 1, with 0 representing total degradation and 1 no degradation. In our application, the degradation is associated to accessibility based on distances to a road network, the closer a location is to a road, the closer $q(\textbf{s})$ is to $1$.\\ \\
The way $q(\textbf{s})$ can be specified is still an open question, and several alternatives are available, depending on the sources of variation in sampling effort that are considered in the model. For example, in the case of moose distribution in Hedmark, $q(\textbf{s})$ could be associated to accessibility to the road system, \citep{doi:10.1002/ecm.1372}, to populated areas and freshwater, \citep{doi:10.1111/ecog.03944}, or land transformation, \citep{doi:10.1111/j.1467-9876.2011.00769.x}. As pointed out in \citet{yuan2017point}, in case $q(\textbf{s})$ is not log-linear, the estimation of the parameters is not part of the latent Gaussian model framework of INLA. Thus, following the half normal detection function in distance sampling, \citep{yuan2017point}, we aim to account for differences in accessibility by making use of the functional form:
\begin{equation}
q(\textbf{s})= \exp(-\zeta \cdot d(\textbf{s})^2 /2 ); \quad \zeta>0
\label{qequation}
\end{equation}
where $\zeta$ is a scale parameter and $d(\textbf{s})$ is the closest distance from location $\textbf{s}$ to the road system. Thus, the model we propose, which accounts for differences in accessibility is:\\ \\
\begin{equation}
\log(\lambda(\textbf{s})q(\textbf{s}))=\textbf{x}^T(\textbf{s})\beta + \omega(\textbf{s}) + \log(q(\textbf{s}))
\label{VSEModel}
\end{equation}
This model requires that the variables that are used to explain $q(\textbf{s})$, in our application distance to the road system, are available at every $\textbf{s} \in D$.

\subsubsection{Prior specification}

The parameter $\nu$ in the Matérn covariance function (\ref{MaternCov}) is fixed to be 1. On the other hand, the interest is put on the spatial range $\rho$ and on $\sigma$ with $\rho$ related to $\kappa$ in (\ref{MaternCov}) through $\rho=\sqrt{8}/\kappa$. These two parameters are specified by making use of PC priors, \citep{doi:10.1080/01621459.2017.1415907}. In this case we set $P(\rho<15)=0.05$ and $P(\sigma>1)=0.05$. It means that under this prior specification a standard deviation greater than 1 is regarded as large, while a spatial range less than 15 is considered unlikely.  The parameters in $\boldsymbol{\beta}$ have Normal prior with mean 0 and precision 0.01. Finally, let $\zeta=\exp(\theta)$. For the hyparameter $\theta$ a Normal prior distribution with mean 1 and precision 0.05 is specified.

\subsection{Inference and computational approach } 

The models introduced in Section 3.1 will be fitted making use of the Integrated Nested Laplace Approximation (INLA), \citep{doi:10.1111/j.1467-9868.2008.00700.x}, the SPDE approach, \citep{doi:10.1111/j.1467-9868.2011.00777.x}, and the approach introduced in \citet{10.1093/biomet/asv064} for fitting spatial point processes.

\subsubsection{The Integrated Nested Laplace Approximation (INLA)}

The traditional approach for performing Bayesian inference for latent Gaussian models is Monte Carlo Markov Chains (MCMC). However, the Integrated Nested Laplace Approximation (INLA), \citep{doi:10.1111/j.1467-9868.2008.00700.x}, has emerged as a reliable alternative, \citep{doi:10.1111/2041-210x.12017,HUMPHREYS2017192,doi:10.1002/ece3.3081}. While MCMC requires considerable time to perform Bayesian inference for complex structures such as those inherent to spatial models, INLA requires less time to do the same task since, unlike MCMC which is simulation based, INLA is a deterministic algorithm, \citep{blangiardo2015spatial}. The aim of INLA is to produce a numerical approximation of the marginal posterior distribution of the parameters and hyperparameters of the model. In addition to its computational benefits, implementing INLA is simple by making use of the R-INLA library.

\subsubsection{The SPDE approach}

An useful and efficient way to represent a continuous spatial process based on a discretely indexed spatial random process is the Stochastic Partial Differential Equation (SPDE) approach, \citep{doi:10.1111/j.1467-9868.2011.00777.x}. This is based on the solution to the SPDE: 
\begin{equation}
(\kappa^2 - \Delta)^\frac{\alpha}{2} (\tau \xi(\mathbf{s}))=\mathbf{W}(\mathbf{s})
\end{equation}
where $\mathbf{s}$ is a vector of locations in $\mathbb{R}^2$, $\Delta$ is the Laplacian. $\nu$, $\kappa>0$ and $\tau>0$ are parameters that represent a control for the smoothness, scale and variance, respectively. $\mathbf{W}(\mathbf{s})$ is a Gaussian spatial white noise process.  The solution for this equation, $\xi(\mathbf{s})$, is a stationary Gaussian Field with Matérn covariance function (\ref{MaternCov}). This solution can be approximated through a basis function representation defined on a triangulation of the spatial domain $D$:
\begin{equation}
\xi(\mathbf{s})=\sum_{g=1}^G \phi_g(\mathbf{s}) \tilde{\xi}_g
\end{equation}
where $G$ is the total number of vertices of the triangulation, $\{\phi_g\}$ is the set of basis functions, and $\{\tilde{\xi}_g\}$ are zero-mean Gaussian distributed weights. This way of representing the Gaussian Random Field has been proven to make more efficient the fitting process. Figure 4a displays the triangulation for the moose distribution example.

\subsubsection{Approach for modeling LGCPs}

The traditional way of fitting point process models is by gridding the space and then modeling the intensity on a discrete number of cells. However, this approach becomes unfeasible and computationally expensive as the number of grids increases. Given that gridding the space also implies approximating the location of the observations, it also represents a waste in information in contexts such as Citizen Science where the locations of the observations are collected with considerable precision. Since a better approximation of the continuous random field is achieved by making the size of the cells as small as possible, lattice-based methods become unfeasible as stressed in \citet{10.1093/biomet/asv064}. The approach there introduced is especially useful in situations with uneven sampling effort since the resolution of the approximation can be locally adapted in those regions with low sampling. Some additional details of this approach are now presented.\\\\
Let $\boldsymbol \omega(\textbf{s})$ be a finite-dimensional continuously specified random field defined as:
\begin{equation}
\boldsymbol \omega(\textbf{s}) = \sum_{i=1}^{n} \omega_i \phi_i(\textbf{s})
\end{equation}
Based on this specification, the likelihood of a LGCP conditional on a realization of $\boldsymbol{\omega}$:
\begin{equation}
\log(\pi(\lambda(\cdot)|\boldsymbol \omega) = |\boldsymbol \omega| - \int_{\boldsymbol \omega} \exp(\boldsymbol \omega(\textbf{s}))d\textbf{s} + \sum_{i=1}^N \boldsymbol{\omega}(s_i) 
\end{equation}
can be approximated by :
\begin{equation}
\log(\pi(\lambda(\cdot)|\boldsymbol \omega)) \approx C - \sum_{i=1}^{p} \tilde{\alpha}_i \exp \bigg \{ \sum_{j=1}^{n} \omega_j \phi_j(\tilde{s}_i) \bigg \} + \sum_{i=1}^{N} \sum_{j=1}^{n} \omega_j \phi_j(s_i)
\label{loglgcp}
\end{equation}
with $\tilde{a}_i$ and $\tilde{s}_i$ a set of deterministic weights and locations that can be obtained from a dual mesh with polygons centered at each node of the mesh. Then, $\tilde{\textbf{s}}=\{\tilde{s}_1,\ldots,\tilde{s}_n\}$ are the nodes of the mesh and $\tilde{\textbf{a}}=\{\tilde{a}_1,\ldots,\tilde{a}_n\}$ the areas of the polygons linked to each centroid. These polygons are constructed by making use of the midpoint rule, \citep{10.1093/biomet/asv064}. The dual mesh for our application is shown in Figure 4b.
\begin{figure}[t]
\begin{multicols}{2}
\center
	\includegraphics[width=0.5\textwidth]{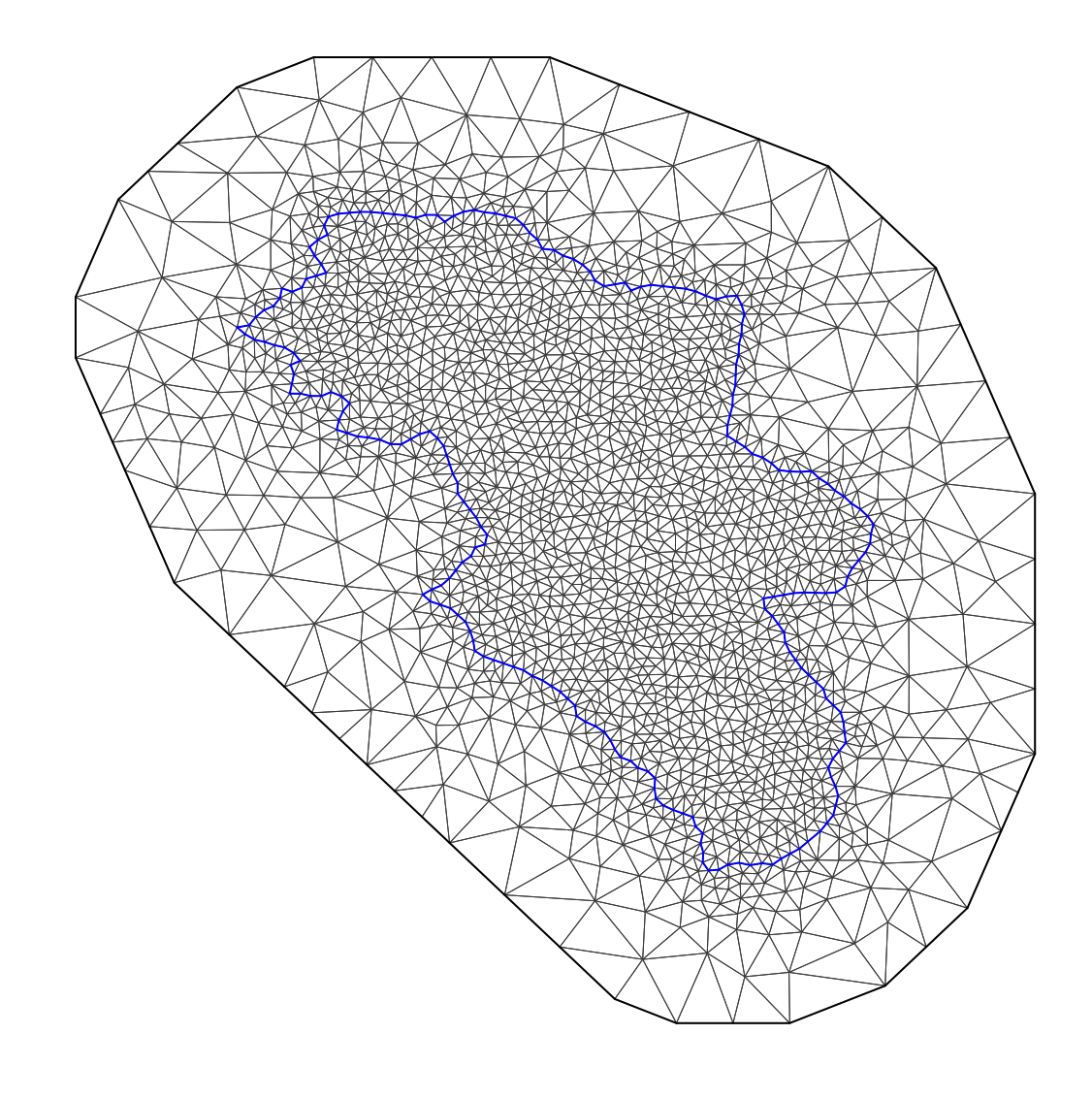} 
	\columnbreak
	\center
\includegraphics[width=0.5\textwidth]{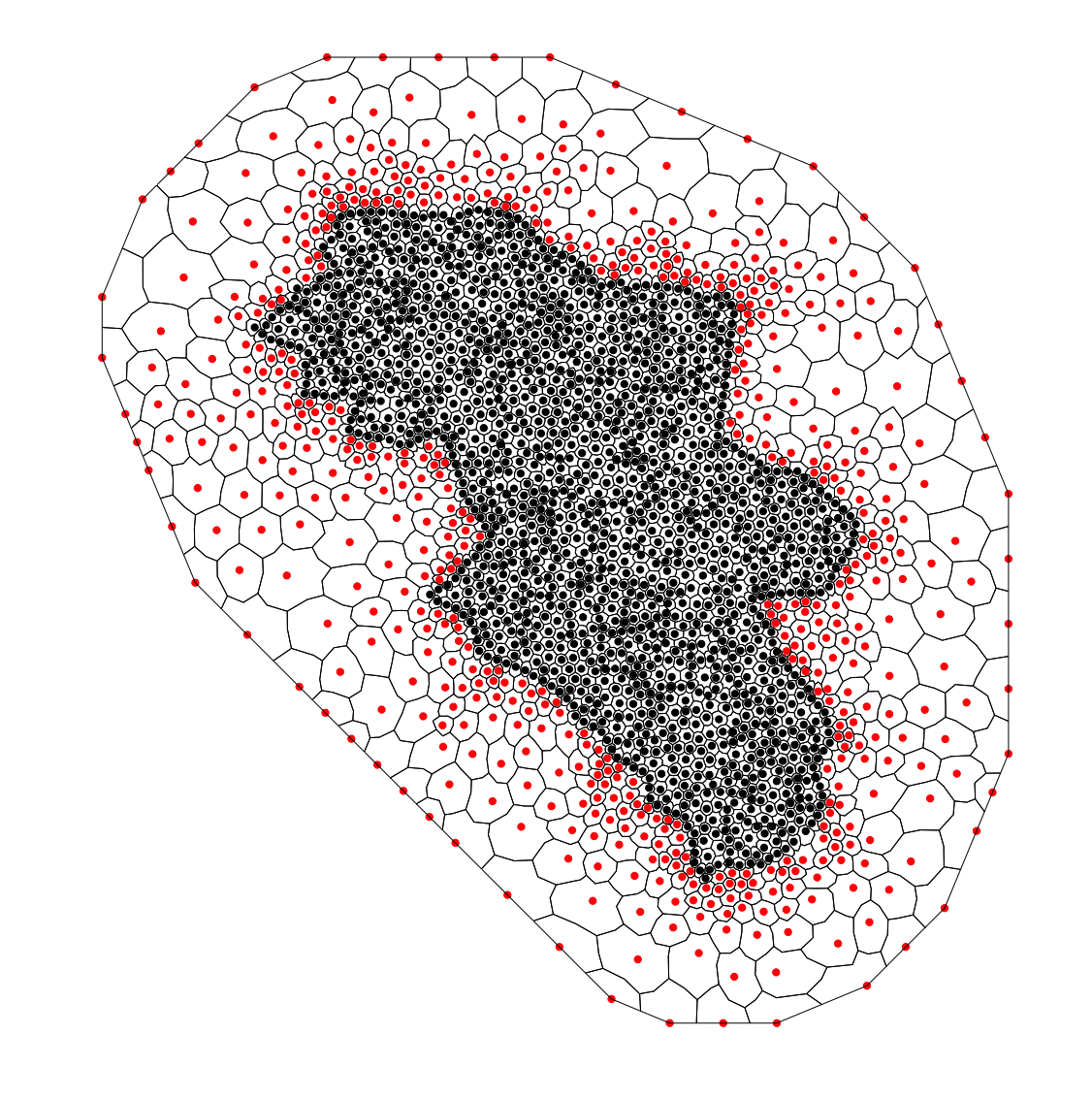} 
\end{multicols}
\begin{minipage}{0.5\textwidth}
	\center
\subcaption{ (a)}
\end{minipage}
\begin{minipage}{0.5\textwidth}
\center
\subcaption{(b)}
\end{minipage}
	\caption{ (a) Triangulation of Hedmark according to the SPDE approach (b) Dual mesh for approximating the likelihood of the LGCP associated to moose distribution in Hedmark. The points are the locations $\tilde{s}_i$ in eq. (\ref{loglgcp}) and the areas of the polygons are the weights $\tilde{a}_i$ in eq. (\ref{loglgcp}).} 
\end{figure}

\subsection{Model assessment}

In order to assess and compare competing models such as the ones we are fitting in upcoming sections, we employ the Deviance Information Criterion (DIC), \citep{doi:10.1111/1467-9868.00353}, the Watanabe-Akaike Information Criterion (WAIC), \citep{Watanabe:2010:AEB:1756006.1953045}, and the logarithm of the pseudo marginal likelihood (LPML).
DIC makes use of the deviance of the model
\begin{equation*}
D(\theta) = -2 \log(p(\textbf{y}|\boldsymbol{\theta}))
\end{equation*}
to compute the posterior mean deviance $\bar{D}=E_{\boldsymbol{\theta}|\textbf{y}}(D(\boldsymbol{\theta}))$. In order to penalize the complexity of the model, the effective number of parameters,
\begin{equation*}
p_D = E_{\boldsymbol{\theta}|\textbf{y}}(D(\boldsymbol{\theta})) - D(E_{\boldsymbol{\theta}|\textbf{y}}(\boldsymbol{\theta})) = \bar{D} - D(\bar{\boldsymbol{\theta}})
\end{equation*}
is added to $\bar{D}$. Thus, 
\begin{equation*}
DIC = \bar{D} + p_D.
\end{equation*}
The Watanabe-Akaike Information Criterion is based on the posterior predictive density, which makes it preferable to the Akaike and the deviance information criteria, since according to \citet{Gelman2014} it averages over the posterior distribution rather than conditioning on a point estimate. It is empirically computed as
\begin{equation*}
-2\bigg[\sum_{i=1}^{n} \log\bigg(\frac{1}{S} \sum_{s=1}^{S} p(y_i|\theta^s) \bigg) + \sum_{i=1}^n V_{s=1}^S ( \log p(y_i|\theta^s)) \bigg]
\end{equation*}
with $\theta^s$ a sample of the posterior distribution and $V_{s=1}^S$ the sample variance\\\\
Another criterion to compare the models is LMPL,defined as:
\begin{equation*}
LPML = \sum_{i=1}^n \log ({CPO}_i)
\end{equation*}
It depends on $CPO_i$, the Conditional Predictive Ordinate at location $i$, \citep{doi:10.1111/j.2517-6161.1990.tb01780.x}, a measure that assesses the model performance by means of leave-one-out cross validation. It is defined as:
\begin{equation*}
CPO_i = p(y_i^*|y_f)
\end{equation*}
with $y_i^*$ the prediction of $y$ at location $i$ and $y_f = y_{-i}$.\\ \\

\section{Simulation Study}

Our simulation study aims to show: i) the implications of not accounting for variations on sampling effort when CS data is modeled and ii) how accounting for at least one source of variation in sampling effort can contribute to improve the inference made about the point process underlying the spatial distribution of a species. In order to do it, we make use of the same region map, the road system in the application, the covariate Solar Radiation (RAD), given its association with the sampling process (82\% of the reports are made in locations whose solar radiation is above the median solar radiation of the entire region) and its negative correlation, $(-0.43)$, with the distance to the road system. Then a zero-mean Gaussian random field with Matérn covariance function is simulated.\\ \\
A point pattern whose intensity depends on RAD is simulated. This is specified as a Log-Gaussian Cox Process, $Y(\textbf{s})$, with log-intensity given by:
\begin{equation}
\log(\lambda(\textbf{s})) = \beta_0 + \beta_1 RAD(\textbf{s}) + \omega(\textbf{s})
\label{Simuequation}
\end{equation}
It is simulated with $\beta_0=-4.25$ and $\beta_1=0.82$. The parameters of the Matérn covariance associated to the zero-mean Gaussian field, $\omega(\textbf{s})$, are assumed to be $\nu=1$, $\kappa \approx \sqrt{8}/\rho = \sqrt{8}/34$, \citep{doi:10.1111/j.1467-9868.2011.00777.x}, with $\rho$ the practical range, and $\sigma^2 = 0.7 $.\\ \\
After simulating the LGCP, we thin the set of sampled points. The probability of keeping a point in the point pattern depends on its distance to the road system, see equation (\ref{qequation}). Figure 5 displays both the map of the distance to the road system in Hedmark as well as the map of the probability of keeping a location in the sample given that it is part of the initially sampled point pattern when $\zeta=16$.
\begin{figure}[t]
	\begin{multicols}{2}
		\includegraphics[width=0.4\textwidth]{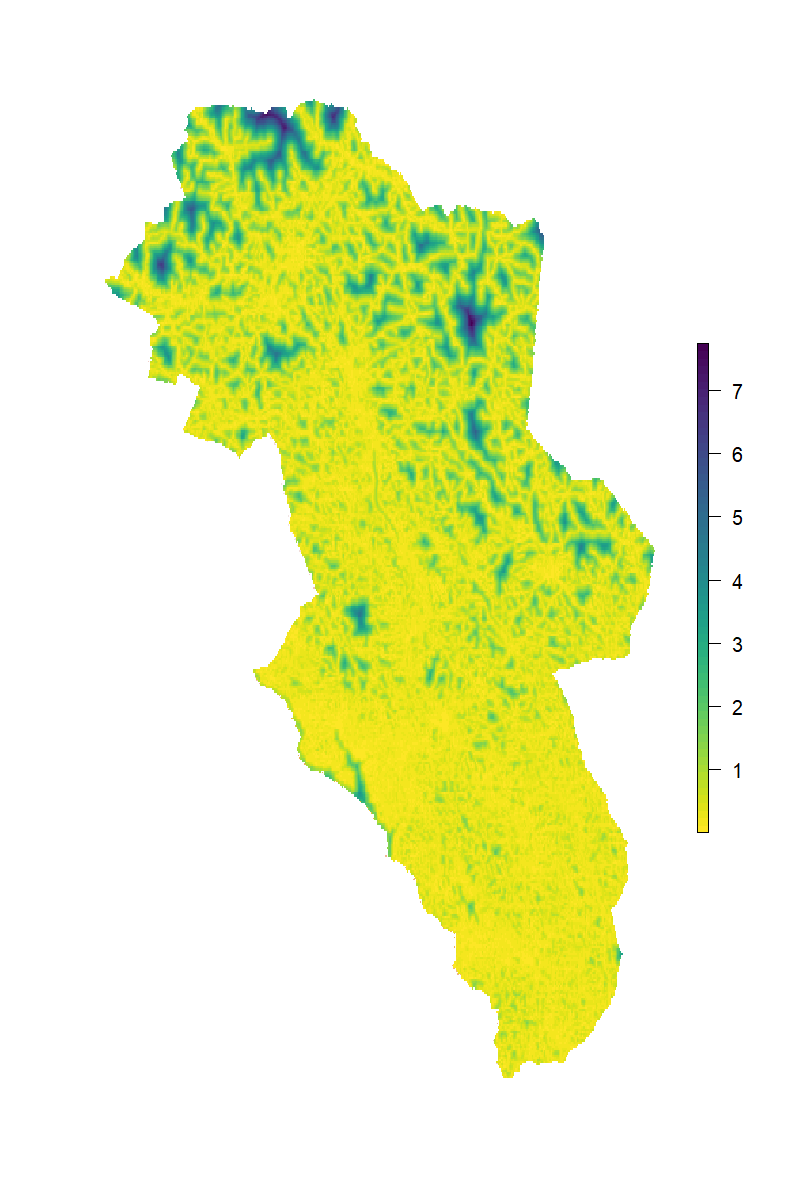} 
		\columnbreak
		\includegraphics[width=0.4\textwidth]{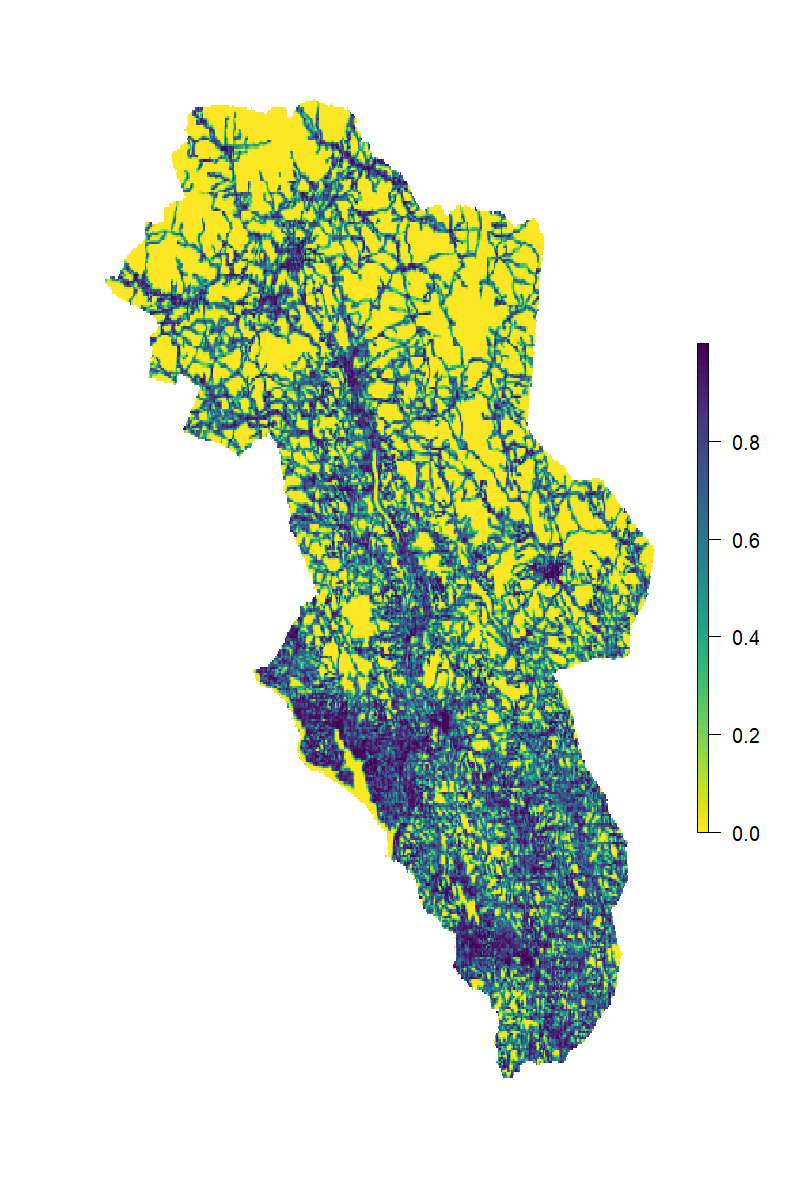} 
	\end{multicols}
	\begin{minipage}{0.5\textwidth}
	\center
\subcaption{ (a)}
\end{minipage}
\begin{minipage}{0.5\textwidth}
\center
\subcaption{(b)}
\end{minipage}
	\caption{ (a) Distance to closest road (in km) in the county of Hedmark, Norway and (b) Access probability when $\zeta=16$ }
\end{figure}

In order to thin the process, we create four scenarios based on the value of $\zeta$ in equation \ref{qequation}: scenario 0, when $\zeta=0$; scenario 1, when $\zeta=1$;
scenario 2, when $\zeta=8$ and scenario 3, when $\zeta=16$. $\zeta=0$ corresponds to the case with no thinning. The other three values of $\zeta$ represent increasing levels of thinning that result in about 12\%, 38\% and 49\% of observations removed, respectively. The process of simulating a LGCP and thinning it according to $\zeta = \{0,1,8,16\}$ is made for 100 different simulated point patterns.\\ \\
To assess the performance of each approach under each scenario, we simulate 100 realizations $\{\theta^p_{jkl}\}, j=,1\ldots,100$, from the posterior distribution of each parameter $\theta$ for point pattern $k =1,\ldots,100$ under scenario $l = 1,2,3,4$. Then, the bias and the Root Mean Square Error (RMSE) for point pattern $k$ in scenario $l$ are computed as:
\begin{align*}
 bias &= \frac{1}{100} \sum_{j=1}^{100} \big(\theta^p_{jkl}-\tilde{\theta} \big)\\ \\
 RMSE &= \large \sqrt{\frac{1}{100} \sum_{j=1}^{100} \big(\theta^p_{jkl}-\tilde{\theta} \big)^2}
\end{align*}
with $\tilde{\theta}$ the actual value of parameter $\theta$.

\section{Results}

\subsection{Simulation Study}
The point patterns obtained under each simulation scenario were fitted using both the naive and the varying sampling effort (VSE) model. Results are now presented. As measures of performance bias, RMSE, introduced in Section 4, and frequentist coverage are used.\\ \\
Remember that under this simulation scenario the intensity when $\zeta=0$ is given by
    \begin{equation*}
\lambda(\textbf{s}) = \beta_0 + \beta_1 RAD(\textbf{s}) + \omega(\textbf{s})
\end{equation*}
with $\omega(\textbf{s})$ depending on the hyperparameters $(\rho,\sigma)$. From this general model, four scenarios were derived: $\zeta=(0,1,8,16)$. As expected, an increment in the thinning parameter, i.e. more accessibility bias, yields loss in accuracy as seen in Figure 6, which displays both the mean bias and the mean RMSE under each scenario for parameter $\beta_1$.\\ \\
\begin{figure}[H]
 		\includegraphics[width=1\textwidth]{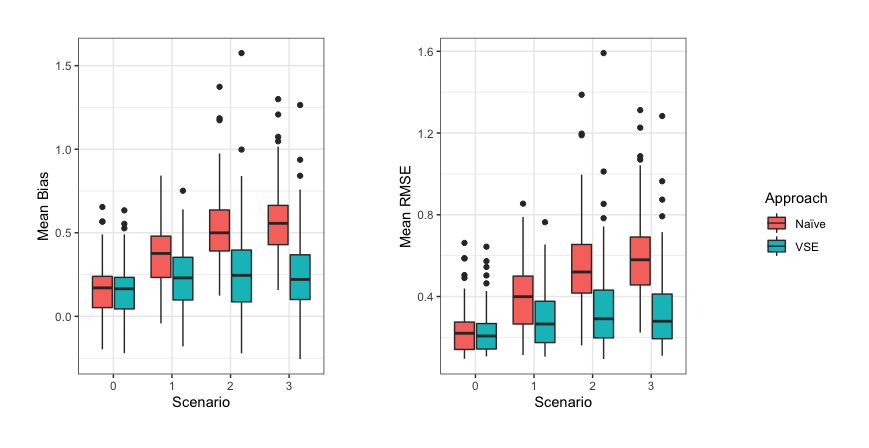} 
 		\caption{Distribution over simulations of the bias and RMSE for $\beta_1$, by scenario and methodological approach  }
 	 \end{figure}
For simulation scenario 0, $\zeta=0$ represents a completely observed point process. Hence, lower mean bias and mean RMSE are obtained in comparison to the other scenarios. When, $\zeta \in \{1,8,16\}$ these two performance indices increase given the deletion of observations. However, it is clear how the corresponding bias and RMSE are smaller under the VSE approach, especially in those scenarios with larger values of $\zeta$. The only parameter for which the bias and RMSE are larger under the VSE approach is $\sigma$. Even though the spatial parameter $\rho$ is less biased under the VSE approach it is clearly overestimated under both approaches.
The spatial variance and the range are the most difficult parameters to estimate and prior distributions that provide more information about these parameters may be useful to improve the accuracy of their estimates, \citep{CAMELETTI2019100353, bakar2015sptimer}.
In similar fashion the estimates of the parameter $\zeta$ become less accurate as less observations are available.\\
% Table generated by Excel2LaTeX from sheet 'Bias'
\begin{table}[t]
  \scriptsize 
  \centering
  \caption{Mean Bias and RMSE for the parameters of the naive and the VSE model under the 4 scenarios simulated. In parenthesis the standard deviation.}
    \begin{tabular}{llcccccccccc}
    \toprule
    Scenario & Approach & \multicolumn{2}{c}{$\beta_0$} & \multicolumn{2}{c}{$\beta_1$} & \multicolumn{2}{c}{$\rho$} & \multicolumn{2}{c}{$\sigma$} & \multicolumn{2}{c}{$\zeta$} \\
\cmidrule{3-12}          &       & Bias  & RMSE  & Bias  & RMSE  & Bias  & RMSE  & Bias  & RMSE  & Bias  & RMSE \\
    \midrule
    \multirow{4}[4]{*}{0} & \multirow{2}[2]{*}{Naive} & 0,160 & 0,274 & 0,158 & 0,231 & 14,823 & 18,471 & -0,038 & 0,158 &       &  \\
          &       & (0,13) & (0,082) & (0,159) & (0,11) & (8,153) & (7,88) & (0,11) & (0,041) &       &  \\
         & \multirow{2}[2]{*}{VSE} & 0,168 & 0,281 & 0,152 & 0,231 & 15,037 & 18,529 & -0,033 & 0,156 & 0,022 & 0,046 \\
          &       & (0,13) & (0,08) & (0,16) & (0,107) & (8,486) & (8,142) & (0,111) & (0,042) & (0,041) & (0,045) \\
\midrule   \multirow{4}[4]{*}{1} & \multirow{2}[2]{*}{Naive} & -0,132 & 0,274 & 0,379 & 0,411 & 13,293 & 17,116 & -0,020 & 0,164 &       &  \\
          &       & (0,153) & (0,09) & (0,185) & (0,169) & (7,71) & (7,443) & (0,124) & (0,058) &       &  \\
        & \multirow{2}[2]{*}{VSE} & 0,151 & 0,281 & 0,241 & 0,303 & 12,994 & 17,052 & -0,038 & 0,167 & -0,085 & 0,280 \\
          &       & (0,159) & (0,092) & (0,191) & (0,148) & (7,944) & (7,512) & (0,126) & (0,059) & (0,213) & (0,099) \\
    \midrule
    \multirow{4}[4]{*}{2} & \multirow{2}[2]{*}{Naive} & -0,633 & 0,670 & 0,538 & 0,564 & 13,362 & 18,102 & -0,006 & 0,180 &       &  \\
          &       & (0,173) & (0,168) & (0,217) & (0,211) & (9,005) & (8,656) & (0,144) & (0,077) &       &  \\
        & \multirow{2}[2]{*}{VSE} & 0,046 & 0,260 & 0,269 & 0,339 & 11,946 & 16,091 & -0,059 & 0,179 & -2,030 & 2,246 \\
          &       & (0,166) & (0,084) & (0,254) & (0,217) & (9,532) & (9,077) & (0,159) & (0,098) & (1,028) & (0,826) \\
    \midrule
    \multirow{4}[4]{*}{3} & \multirow{2}[2]{*}{Naive} & -0,877 & 0,905 & 0,569 & 0,596 & 13,717 & 19,087 & 0,002 & 0,192 &       &  \\
          &       & (0,188) & (0,189) & (0,214) & (0,209) & (11,585) & (11,436) & (0,152) & (0,077) &       &  \\
        & \multirow{2}[2]{*}{VSE} & -0,012 & 0,265 & 0,244 & 0,329 & 10,511 & 13,749 & -0,090 & 0,188 & -4,366 & 4,534 \\
          &       & (0,18) & (0,087) & (0,243) & (0,193) & (10,269) & (9,219) & (0,17) & (0,096) & (2,099) & (1,923) \\
    \bottomrule
    \end{tabular}%
  \label{tab:addlabel}%
\end{table}%
As an additional comparison measure we used the frequentist coverage of the equal-tailed $100(1-\alpha)\%$ Bayesian credible intervals for each parameter. The coverage of the spatial parameters is high. In the case of the spatial range it is due to the large variability in its posterior densities. It is worth noting that smaller coverages are obtained for the parameters involved in the mean of the intensity, $\beta_0$ and $\beta_1$. They achieve lower values in spite of having bias and RMSE closer to zero. The decrease in the coverage of $\zeta$ is associated to the increasing bias and RMSE as the thinning parameter increases.\\ \\
% Table generated by Excel2LaTeX from sheet 'nvcoverage'
\begin{table}[htbp]
  \centering
  \caption{Frequentist coverage for the parameters of the naive model under the 4 simulation scenarios}
    \begin{tabular}{lrrrr}
    \toprule
    \multicolumn{1}{c}{Scenario} & \multicolumn{1}{c}{$\beta_0$} & \multicolumn{1}{c}{$\beta_1$} & \multicolumn{1}{c}{$\rho$} & \multicolumn{1}{c}{$\sigma$} \\
    \midrule
    0     & 0,92 (0,79) & 0,79 (0,49) & 0,79 (39,24) & 0,92 (0,44) \\
    1     & 0,96 (0,79) & 0,29 (0,55) & 0,88 (38,92) & 0,92 (0,46) \\
    2     & 0,1 (0,83) & 0,11 (0,63) & 0,9 (42,9) & 0,91 (0,5)  \\
    3     & 0,01 (0,86) & 0,09 (0,67) & 0,93 (46,41) & 0,91 (0,53)  \\
    \bottomrule
    \end{tabular}%
  \label{tab:addlabel}%
\end{table}%

% Table generated by Excel2LaTeX from sheet 'vsecoverage'
\begin{table}[htbp]
  \centering
  \caption{Frequentist coverage for the parameters of the VSE model under the 4 simulation scenarios}
    \begin{tabular}{lrrrrr}
    \toprule
    \multicolumn{1}{c}{Scenario} & \multicolumn{1}{c}{$\beta_0$} & \multicolumn{1}{c}{$\beta_1$} & \multicolumn{1}{c}{$\rho$} & \multicolumn{1}{c}{$\sigma$} & \multicolumn{1}{c}{$\zeta$} \\ \midrule
    0     & 0,94 (0,77) & 0,81 (0,49) & 0,75 (38,80) & 0,93 (0,43) & 0,00 (0,02) \\
    1     & 0,89 (0,77) & 0,6 (0,54) & 0,9 (39,07) & 0,88 (0,45) & 0,83 (0,72) \\
    2     & 0,97 (0,81) & 0,62 (0,61) & 0,82 (37,08) & 0,84 (0,42) & 0,24 (2,78) \\
    3     & 0,96 (0,81) & 0,65 (0,64) & 0,63 (26,53) & 0,55 (0,31) & 0,09 (2,86) \\
    \bottomrule
    \end{tabular}%
  \label{tab:addlabel}%
\end{table}%
The model comparison methods based on the deviance and on the predictive distribution as the ones introduced in Section 3 are used to compare the models fitted under both approaches. In the scenario with $\zeta=0$ the naive model is the true model and, as expected, it performed better than the VSE approach in about 60\% of the simulated point patterns. As expected, this situation changes as the thinning parameter increases. Thus the VSE model performs better than the naive one for about all the simulated datasets according to the DIC and the WAIC, while it is better according to the CPO for 90\% of the patterns simulated.\\ \\
If there is evidence of stronger variation in sampling effort due to difference in accessibility, an informative prior distribution with larger mean would reduce both the bias and the RMSE for this parameter. In order to demonstrate it, we change the prior specification of the hyperparameter $\theta$, associated to $\zeta$, formerly of mean 1 and precision 0.05, to have precision 10 and mean 3.5 and 5 for scenarios 2 and 3, respectively. Mean bias, mean RMSE and the frequentist coverages for these scenarios are presented in tables 4 and 5. 
% Table generated by Excel2LaTeX from sheet 'Bias'
\begin{table}[htbp]
  \scriptsize 
  \centering
  \caption{Mean Bias and RMSE for the parameters of the VSE model under scenarios 2 and 3 with the new prior specification.}
    \begin{tabular}{lcccccccccc}
    \toprule
    Scenario  & \multicolumn{2}{c}{$\beta_0$} & \multicolumn{2}{c}{$\beta_1$} & \multicolumn{2}{c}{$\rho$} & \multicolumn{2}{c}{$\sigma$} & \multicolumn{2}{c}{$\zeta$} \\
\cmidrule{2-11}            

& Bias  & RMSE  & Bias  & RMSE  & Bias  & RMSE  & Bias  & RMSE  & Bias  & RMSE \\
    \midrule
     \multirow{2}[2]{*}{2}   & 0,104 & 0,288 & 0,304 & 0,380 & 9,826 & 14,069 & -0,040 & 0,199 & -0,347 & 1,176 \\
                & (0,240) & (0,173) & (0,742) & (0,724) & (9,450) & (8,496) & (0,333) & (0,287) & (1,020) & (0,457) \\
    \midrule
   \multirow{2}[2]{*}{3}     & 0,098 & 0,276 & 0,236 & 0,327 & 10,411 & 13,970 & -0,064 & 0,177 & -0,199 & 2,138 \\
                & (0,173) & (0,095) & (0,260) & (0,210) & (11,921) & (11,159) & (0,182) & (0,110) & (2,219) & (0,984) \\
    \bottomrule
    \end{tabular}%
  \label{tab:addlabel}%
\end{table}%
 
 In comparison to the previous results, the bias and RMSE for both scenarios are considerably reduced for parameter $\zeta$. For the other parameters these indices suffer minor changes.

% Table generated by Excel2LaTeX from sheet 'vsecoverage'
\begin{table}[htbp]
  \centering
  \caption{Frequentist coverage for the parameters of the VSE model under scenarios 2 and 3 with the new prior specification.}
    \begin{tabular}{lrrrrr}
    \toprule
    \multicolumn{1}{c}{Scenario} & \multicolumn{1}{c}{$\beta_0$} & \multicolumn{1}{c}{$\beta_1$} & \multicolumn{1}{c}{$\rho$} & \multicolumn{1}{c}{$\sigma$} & \multicolumn{1}{c}{$\zeta$} \\    \midrule
    2     & 0,92 (0,79) & 0,62 (0,61) & 0,79 (31,89) & 0,73 (0,36) & 0,72 (2,49) \\
    3     & 0,95 (0,82) & 0,68 (0,65) & 0,59 (24,38) & 0,55 (0,28) & 0,31 (2,27) \\
    \bottomrule
    \end{tabular}%
  \label{tab:addlabel}%
\end{table}%

As expected, the frequentist coverage for $\zeta$ in both scenarios improves significatively, changing from 0,24 to 0,72 when $\zeta=8$ and from 0,09 to 0,31 when $\zeta=16$.

\subsection{Results for moose distribution in Hedmark application}

Both, the naive approach and the VSE approach, introduced in Section 3, are used to model the intensity of moose presence. Table 6 reports the posterior summaries of the parameters involved in each of these models. TRI is negatively related to the intensity, while RAD has positive association with it.  The variability and range associated to the Gaussian field have right skewed posterior distributions based on their posterior medians and means. There is a difference in the posterior mean of RAD coefficient between the models. It is larger when differences in accessibility are not considered in the model. In addition to it, both parameters associated to the Matérn Gaussian field have lower posterior medians under the VSE model.\\ \\
\begin{table}[t]
	\centering
	\caption{Posterior summaries of the parameters of the naive and the VSE model for the moose presence data in Hedmark, Norway}
	\begin{tabular}{l|rrrrr|rrrrr}
		\cmidrule{2-11}    \multicolumn{1}{r}{} & \multicolumn{10}{c}{Model} \\
		\cmidrule{2-11}    \multicolumn{1}{r}{} & \multicolumn{5}{c|}{Naive}            & \multicolumn{5}{c}{VSE} \\
		\multicolumn{1}{c|}{Parameter} & \multicolumn{1}{c}{Mean} & \multicolumn{1}{c}{Sd} & \multicolumn{1}{c}{0.025q} & \multicolumn{1}{c}{0.50q} & \multicolumn{1}{c|}{0.975q} & \multicolumn{1}{c}{Mean} & \multicolumn{1}{c}{Sd} & \multicolumn{1}{c}{0.025q} & \multicolumn{1}{c}{0.50q} & \multicolumn{1}{c}{0.975q} \\
		\midrule
		Intercept & -4.86 & 0.23  & -5.32  & -4.87 & -4.41 & -4.56 & 0.22  & -4.99 & -4.56 & -4.13 \\
		TRI   & -0.20 & 0.08  & -0.36 & -0.20 & -0.04 & -0.20 & 0.08  & -0.35 & -0.20 & -0.04 \\
		RAD   & 1.01  & 0.19  & 0.64  & 1.01  & 1.38  & 0.73  & 0.18  & 0.37  & 0.73  & 1.10 \\
		zeta  & \multicolumn{1}{c}{-} & \multicolumn{1}{c}{-} & \multicolumn{1}{c}{-} & \multicolumn{1}{c}{-} & \multicolumn{1}{c|}{-} & 0.88  & 0.21  & 0.52  & 0.86  & 1.32 \\
		rho (Km) & 39.78 & 9.06  & 26.13 & 38.32 & 61.37 & 37.84 & 7.64  & 25.81 & 36.79  & 55.63 \\
		sigma & 1.12 & 0.16   & 0.85  & 1.10  & 1.48  & 1.05  & 0.14   & 0.81  & 1.04  & 1.35 \\
		\bottomrule
	\end{tabular}%
	\label{tab:addlabel}%
\end{table}%
The parameter $\zeta$ with posterior median 0.86 indicates that the observed point pattern is a thinned version of the real one. Figure 7 shows the estimated relation between distance (in kilometers) to the road system and $q(\textbf{s})$. According to it a point located more than 3 kilometers away from the road system can be regarded as inaccessible for citizen scientists.\\ \\
\begin{figure}[H]
	\center
		\includegraphics[width=0.8\textwidth]{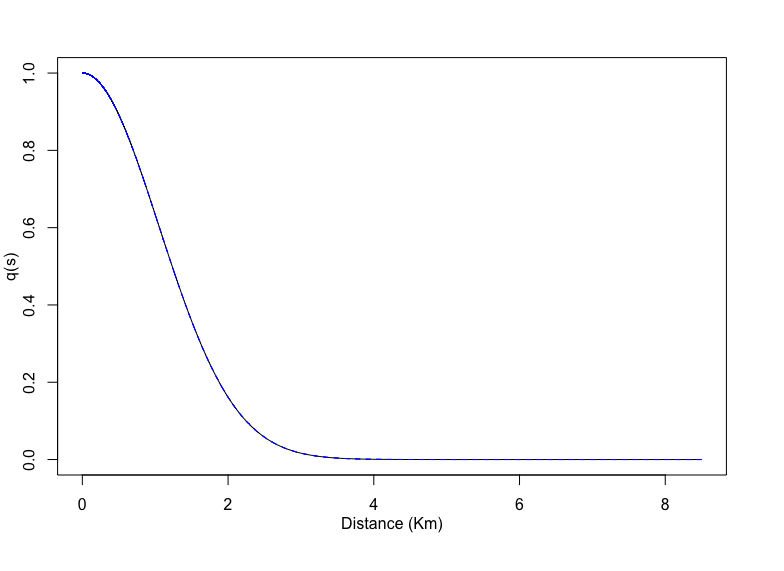} 
		\caption{Estimated relation between distance to the road system, in kilometers, and the probability of having access to location $\textbf{s}$}
		\end{figure}
Figure 8 displays the map of the differences, in log scale, of both the predicted intensity and the standard error associated to the prediction under both approaches. The largest differences occur in zones that are less accessible, which present larger predicted median intensity under the VSE approach than under the naive one. For the zones that are more observed, accounting for differences in accessibility does not affect the predicted median intensity.\\ \\
On the northern side of Hedmark there are zones that are more uncertain. This is expected given that these are zones that are not as accessible as others and have been less sampled. In many location of the northern side of Hedmark the uncertainty is smaller under the VSE approach, while there are not differences in uncertainty in most of the southern half of the region, which is characterized for being the most accessible and sampled part of Hedmark.\\ \\
\begin{figure}[t]
\begin{multicols}{2}
	\includegraphics[width=0.5\textwidth]{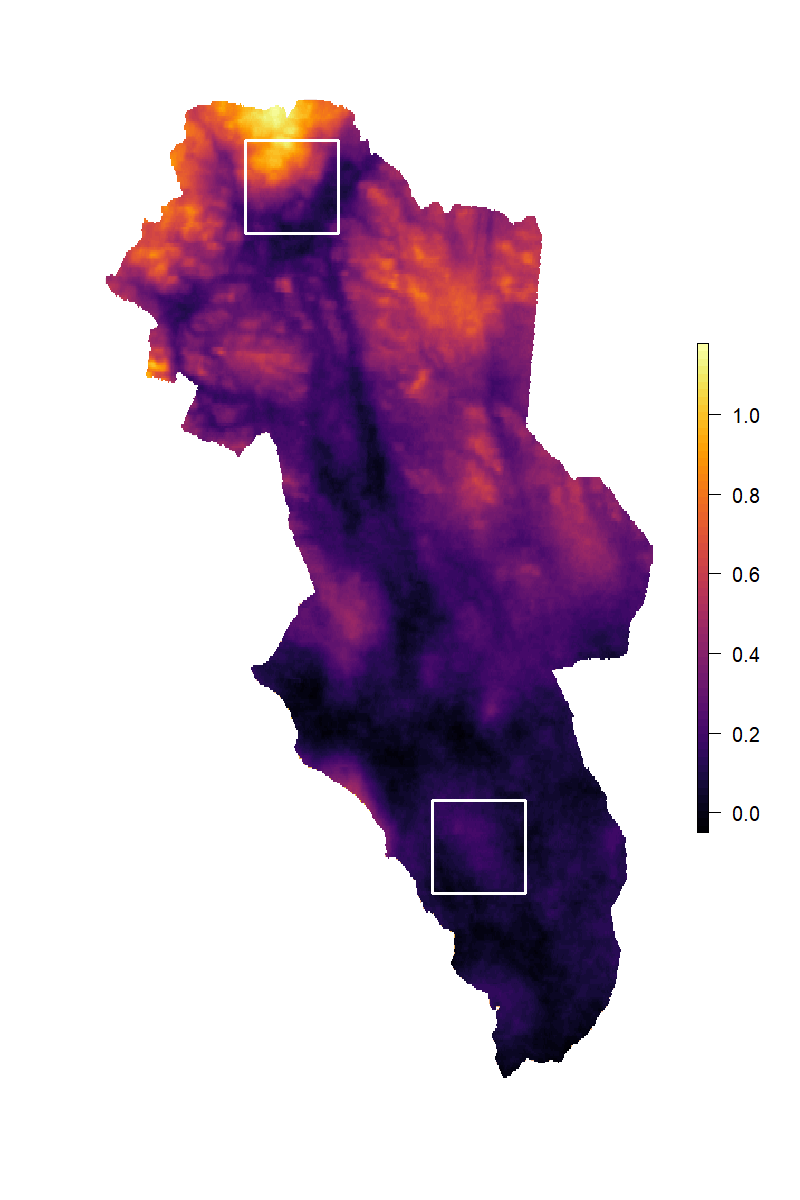} 
	\columnbreak
	\includegraphics[width=0.5\textwidth]{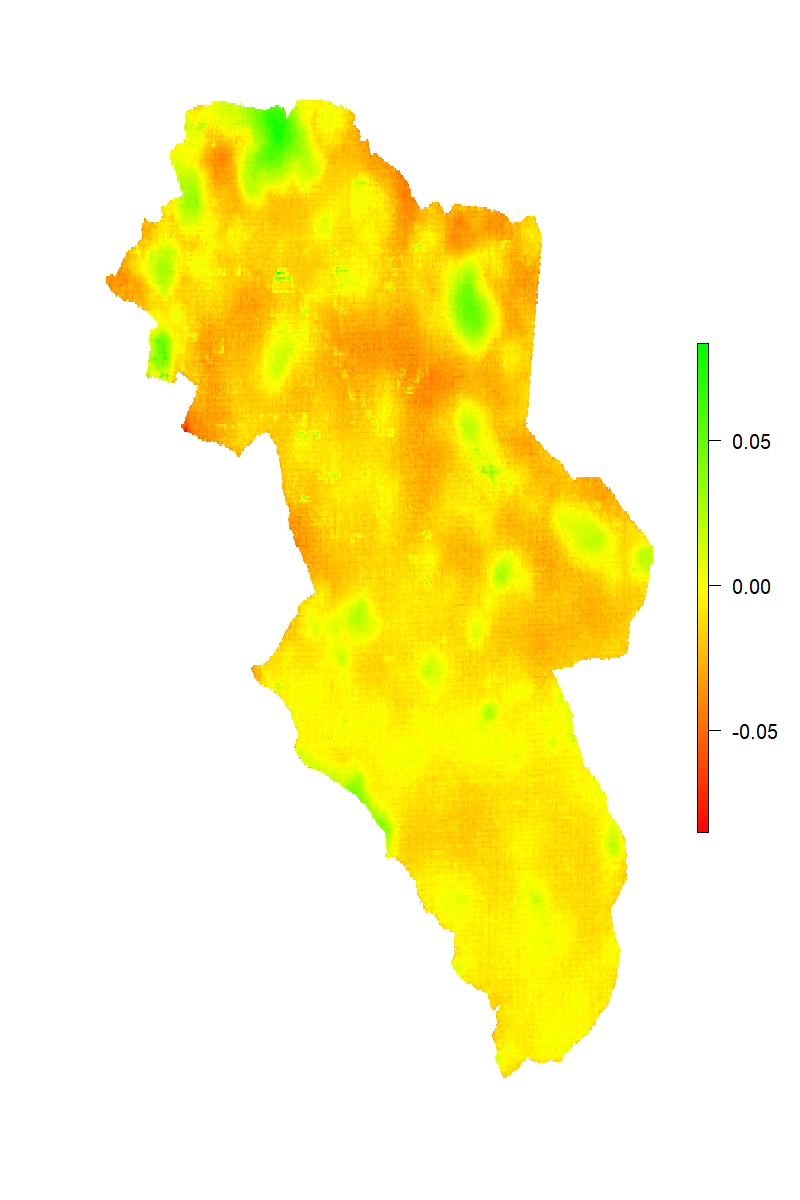} 
\end{multicols}
\begin{minipage}{0.5\textwidth}
	\center
\subcaption{ (a)}
\end{minipage}
\begin{minipage}{0.5\textwidth}
\center
\subcaption{(b)}
\end{minipage}
	\caption{ (a) Differences in predicted median intensity and (b) differences in standard error of the predicted median intensity obtained through the VSE model and the naive model. In (a) the two squares represent the zones that are focused in figures 9 (south of Hedmark) and 10 (north of Hedmark)}
\end{figure}
Both approaches are compared by making use of the DIC, the WAIC and the LPML. Table 7 introduces the value of each criteria under each approach.\\ \\
\begin{table}[H]
	\centering
	\caption{Comparison criteria for the naive and VSE model fitted to moose location reports}
	\begin{tabular}{l|rr}
		\cmidrule{2-3}    \multicolumn{1}{r}{} & \multicolumn{2}{c}{Model} \\
		\cmidrule{2-3}    \multicolumn{1}{r}{} & \multicolumn{1}{c}{Naive} & \multicolumn{1}{c}{VSE} \\
		\midrule
		DIC   & 4377.51 & 4345.38 \\
		WAIC  & 4505.38 & 4473.65 \\
		$\sum \log(CPO_i)$ & -2468.00 & -2446.98 \\
		\bottomrule
	\end{tabular}%
	\label{tab:addlabel}%
\end{table}%
For this particular example the results indicate that accounting for variation in sampling effort represents an improvement in terms of goodness of fit since both the DIC and WAIC take smaller, and the LPML larger values for the VSE model.\\ \\
Now we will focus on two specific zones of Hedmark to see with more detail how the prediction and its associated uncertainty vary between the two approaches. The two zones are bounded by a 30km $\times$30km square and are highlighted in Figure 8. The first zone is located on the southern half of Hedmark between Kongsvinger and Hamar. It is accessible only through service roads, which are not as visited as the main roads of the region, while the second square corresponds to one of the most distant zones of the region, which is located on the northern border of Hedmark. For zone 1 the predicted median intensity and its associated standard error under both approaches is displayed in Figure 9. The predicted median intensity under both approaches is similar as well as the associated uncertainties. Given that the zone is regarded as highly accessible, considerable differences are not expected. For zone 2 the VSE approach increases the intensity in most locations. In terms of uncertainty both approaches produce similar results. However, it gets larger in some few zones under the VSE model, see Figure 10. 

\begin{figure}[H]
	\center
		\includegraphics[width=0.5\textwidth]{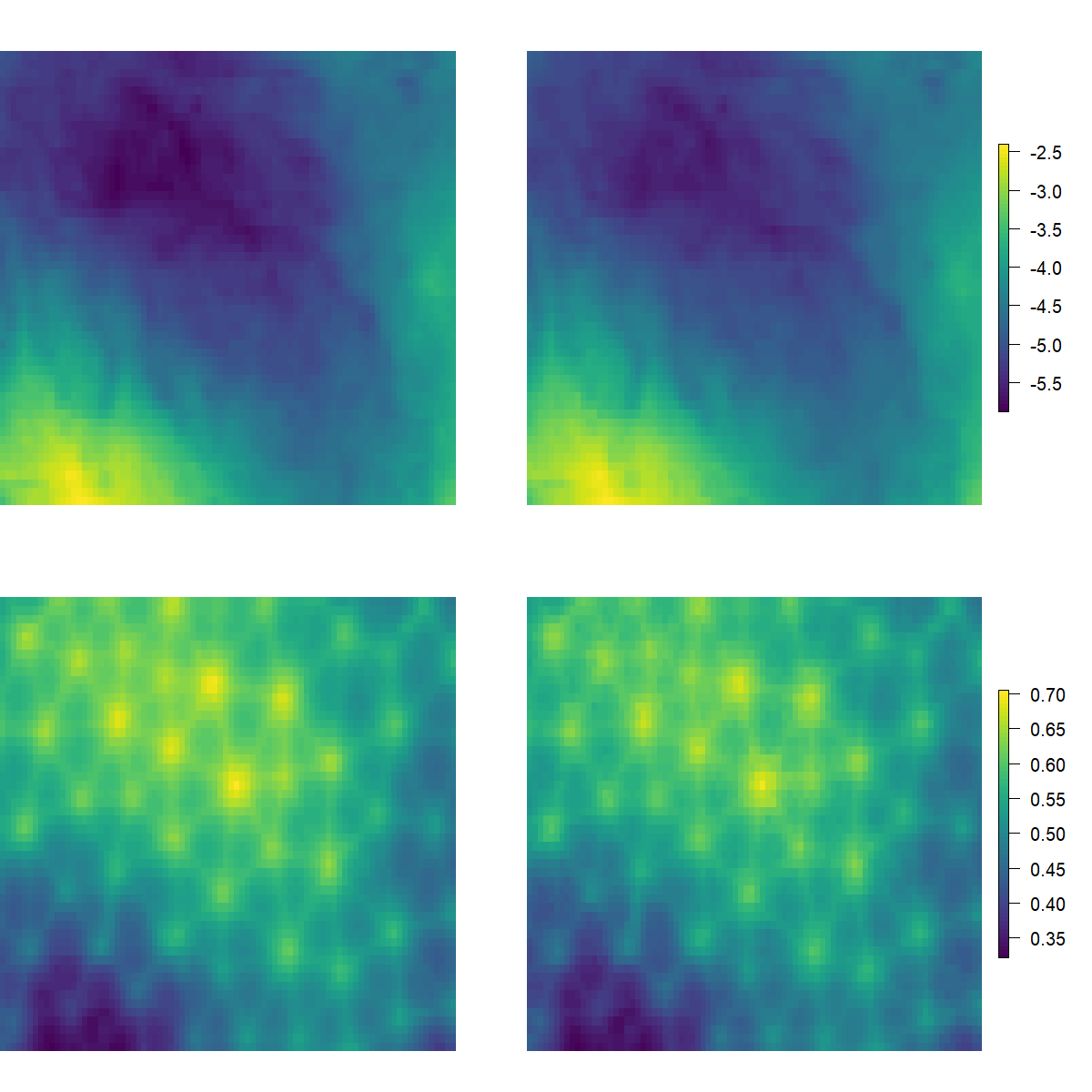} 
		\caption{Predicted median intensity (top) and associated standard error (bottom) for the naive model (left) and the VSE model (right) in zone 1 }
		\end{figure}

		\begin{figure}[H]
	\center
		\includegraphics[width=0.5\textwidth]{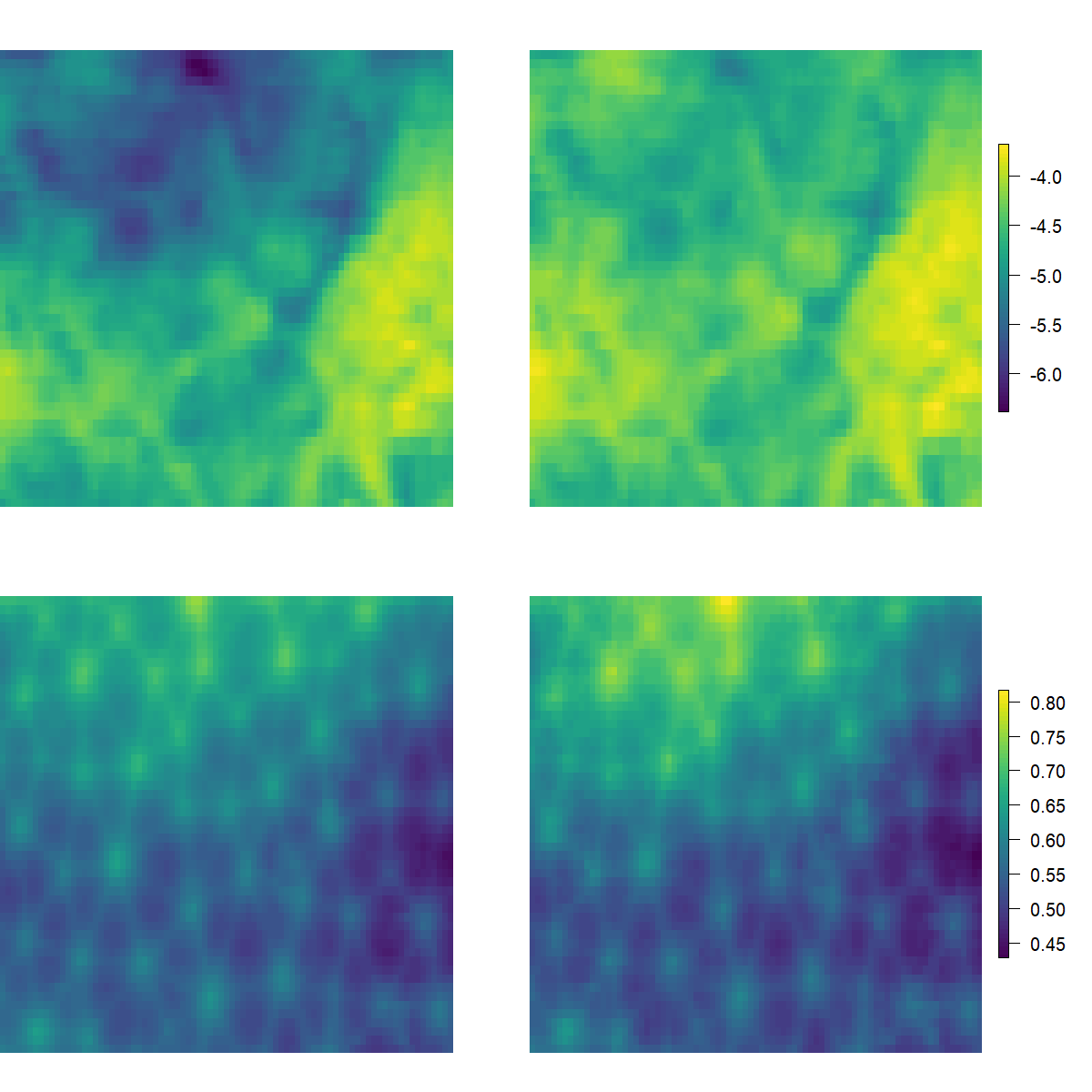} 
		\caption{Predicted median intensity (top) and associated standard error (bottom) for the naive model (left) and the VSE model (right) in zone 2 }
		
		\end{figure}

\section{Discussion and conclusions}
The main goal of this paper was to highlight the importance of accounting for sources of variation in spatial sampling effort for CS data. A Bayesian spatial model that accounts for variation in sampling effort models by including  proxies for external processes that degrade the intensity of the point process has been introduced.\\ \\
This paper focused on differences in accessibility across space. In the simulation study performed in Section 4, we created scenarios where the only source of degradation for the actual point pattern was the distance to the nearest road. The functional form used to link it to the intensity of the point pattern is the half-normal function, \citet{yuan2017point}, characteristic of distance sampling. The aim of ecological studies is often to learn about the effect of covariates. In the moose case study we were interested in exploring the effect of solar radiation and ruggedness on the spatial distribution of moose. The results of both the simulation study and the real data application suggest that in situations with some evidence of uneven sampling effort accounting for differences in accessibility improves performance indices, such as bias and RMSE, and model selection indices, such as DIC, WAIC and LMPL. Furthermore, differences in the covariates posterior summaries in the simulation study showed that in cases with sampling biases the effect of an explanatory may be incorrectly estimated if they are not considered in the model. \\ \\
In our case study we focused on two zones of Hedmark. The large difference in intensity between the VSE approach and naive approach in Zone 2 shows that accounting for differences in accessibility of sampling locations can improve modelling performance when using biased CS data. The differences indicate a need for increased sampling effort in this region, marking the area around Forollhogna national park. This area is one of the few mountainous areas in Norway with relatively gentle slopes and is therefore called the "friendly mountains". Moose occasionally passes through this area, however, only few CS observations have been made so far which might partly be due to a low accessibility and therefore low CS activity.
In contrast, the road network in zone 1 is rather dense. Therefore, the values of $q(\textbf{s}$ are estimated to be relatively high and the model assumes high CS activity in this area. However, the road network here is mainly composed of service roads and small tracks. Therefore, no CS observations of moose in this area might be a result of a low visiting rate of people rather than moose being absent. However, we only accounted for differences in accessibility of sampling locations in space, therefore, the habitat is predicted to be not suitable, which seems to be wrong from an ecological perspective. Accounting for differences in visits of sampling locations in time, for instance by using spatially refined information on type of road or population data could further increase modelling performance. The results highlight, that not accessibility (e.g. roads) are important features for quantifying preferential sampling in CS data, but also how frequent sampling sites are being visited. Small service roads and hiking tracks are likely to have a lower turnover of visiting people than larger roads, and hence, CS more frequently register observations close to larger roads than close to small and remote roads. 
\\ \\
An important part of the VSE model is the parameter $\zeta$, which is necessary to determine to what extent the differences in accessibility affect the observed process. The quality of this estimate was, not surprisingly, proven to improve when more informed prior knowledge is available and endorsed. Alternate specifications of the effect $q(\textbf{s})$ making use of the SPDE approach can be used as in \citet{yuan2017point}. The prior specification of the parameters that are part of the spatial Gaussian field $\omega(\textbf{s})$ is a complex task in spatial statistics. In this paper PC priors were used as a way to incorporate prior knowledge about these parameters in a straightforward way. Alternative prior specifications using PC priors are introduced in \citet{doi:10.1111/rssc.12321}.\\ \\
The VSE model is a first step for modeling CS data in a way that accounts for its inherent sources of bias. More effort is required for e.g.  extending the sampling effort model to more quantities (e.g. cell phone coverage or geographical parameters). Extending the VSE to more spices would be an interesting approach to learn more about citizen science sampling effort in general.

\section*{Acknowledgements}
This paper is part of the NTNU digital transformation project Transforming Citizen Science for Biodiversity.

\clearpage 
\bibliography{RefMoosePaper}

%	

	%--/Paper--
	
\end{document}